\documentstyle[prl,twocolumn,aps]{revtex}
\input{epsf}
\begin{document}
\draft
\twocolumn[\hsize\textwidth\columnwidth\hsize\csname @twocolumnfalse\endcsname
\title{Fast Ground-State Reordering from Off-Diagonal Fluctuations of 
Interaction}

\author{Philippe Jacquod}

\address {Department of Applied Physics, P. O. Box 208284, Yale University,
New Haven, CT 06520-8284, U.S.A.\\
Instituut-Lorentz, Universiteit Leiden
P.O. Box 9506, 2300 RA Leiden, The Netherlands}

\date{\today}

\maketitle
\begin{abstract}
It is shown that off-diagonal fluctuations of interaction lead to
the breakdown of perturbation theory in the immediate vicinity of the
ground-state of a fermionic system 
at a rate inversely proportional to the number of considered particles.
This rate is parametrically much faster than previously expected. 
It is shown that this follows from the coherent addition
of many small second order contributions to the energy of low-lying 
levels.
\end{abstract}
\pacs{PACS numbers : 73.23.-b, 71.10.-w, 71.24.+q, 75.10.Lp}
]

Random interactions naturally occur in various fields of physics.
Assuming the symmetry of the probability distribution of the one-body
Hamiltonian under unitary basis transformation
in Hilbert space directly results in the randomness 
and a similar symmetry for the distribution of the Interaction 
Matrix Elements (IME). Typically, this symmetry is satisfied up
to an energy scale $E_c$ (the Thouless energy) so that models
with random interactions may be expected to describe 
interacting systems 
provided one restricts oneself to energy scales of an order of magnitude
given by $E_c$ \cite{altshklo,nuketh}. In condensed-matter
systems, this corresponds 
to one-body excitation energies inside a window of 
width $E_{c} = g \Delta$ around the Fermi energy, hence containing
a finite number of levels given by the conductance $g$ ($\Delta$
is the mean level spacing). 

Random Interaction Models (RIM) first appeared in nuclear physics
and early investigations focused on the deviations from Random Matrix 
Theory introduced by considering interactions of finite order $k \ll 
n$ ($n$ is the number of particles, for $k=n$ one recovers the
Dyson ensembles of random matrices) \cite{french}. The complexity of the 
excited spectrum of rare-earth atoms 
motivated the introduction of RIM in atomic physics \cite{flam}, and RIM
reproduced statistical features obtained in nuclear shell model
calculations \cite{zelev}.
Last but not least, RIM were used in condensed matter physics to study 
the quasiparticle lifetime \cite{agkl,mirlin} and the conductance peak spacings
and heights distributions \cite{yoram} in quantum dots in the Coulomb blockade 
regime. More recently a lot of attention has been devoted to the
properties of the ground-state of RIM \cite{spings,pert,nukespin} and 
several spectral 
features presented in these latter 
investigations indicate that RIM are realistic down to very low excitation
energies. It is the purpose of the present letter to investigate 
in detail the ground-state 
structure of systems of few ($n$) fermions as the relative strength 
$U/\Delta$ of
fluctuations of interaction increases. The main result of these investigations
is a breakdown of perturbation theory for the ground-state at a
critical value $U_c(n) \sim \Delta/n$. 
In particular, $U_{c}$ is parametrically smaller by a factor $1/n$ 
than previously suggested \cite{sven}. 
Once this threshold is reached, the
ground-state starts to be composed of an exponentially large number
of Slater determinants, an occurrence which
goes far beyond a simple renumbering
of one-particle orbitals. We will see that this is due to the 
coherent addition of large numbers of small second order contributions 
for perturbation theory in the immediate vicinity of the 
ground-state.

We investigate a model for spin-$1/2$ fermions

\begin{eqnarray}\label{hamiltonian}
H & = & \sum \epsilon_{\alpha}
c^{\dagger}_{\alpha,s} c_{\alpha,s}
+\sum
  U_{\alpha,\beta}^{\gamma,\delta}
c^{\dagger}_{\alpha,s} c^{\dagger}_{\beta,s'} c_{\delta,s'}
c_{\gamma,s}
\end{eqnarray}

\noindent The 
greek indices label $m/2$ different one-body energies which are distributed as
$\epsilon_{\alpha} \in [0;m/2]$ so that the mean spacing
between spin-degenerate levels is $\Delta \equiv 1$ and  
$s^{(')} = \uparrow,\downarrow$ are spin indices. The Hamiltonian 
is Spin Rotational Symmetric (SRS): the interaction
commutes with both the total spin $|\vec{S}|$ and its projection
$S_{z}$. The corresponding eigenvalues $\sigma$ and
$\sigma_{z}$ are good quantum numbers and
the Hamiltonian matrix acquires a block structure where blocks are 
labelled by $\sigma_z$ and $\sigma$. From now on we consider only the 
$\sigma=\sigma_z=0$ (1/2) block for $n$ even (odd). The results and arguments 
presented below can however be generalized to higher spin blocks or models 
of spinless fermions.

The Hamiltonian (1) can be viewed as a model of interacting
fermions expressed in the basis of Slater determinants constructed
from the set of one-particle eigenfunctions $\psi_{\alpha}$ of $H_0 \equiv
\sum \epsilon_{\alpha} c^{\dagger}_{\alpha,s} c_{\alpha,s}$. 
In this basis, the IME are given by
$ U_{\alpha,\beta}^{\gamma,\delta} = \sum_{i,j}
U(i-j) \psi_{\alpha}^*(i)  \psi_{\beta}^*(j)
  \psi_{\gamma}(i)  \psi_{\delta}(j)$, where
$U(i-j)$ is the interaction potential in real space; $i$ and $j$ label 
sites on a lattice. Assuming that the presence of disorder or chaotic
boundary scattering 
confers a random character to the
$\psi_{\alpha}$'s directly leads to fluctuations
of the $ U_{\alpha,\beta}^{\gamma,\delta}$ around their average value.
Only diagonal matrix elements have a nonzero average
leading to mean-field charge-charge, spin-spin and BCS
interactions \cite{pert} which we will neglect 
as they have at most a marginal influence on the object of 
interest here. We therefore take the IME in (1) to 
be randomly distributed inside a zero-centred Gaussian distribution 
of width $U$, 
$ P(U_{\alpha,\beta}^{\gamma,\delta}) \propto
e^{-(U_{\alpha,\beta}^{\gamma,\delta})^2/2U^2}$. 

We start with a perturbative calculation of the ground-state energy
up to the second order in the small
parameter $U/\Delta \ll 1$. The first order correction vanishes on 
average, and the second order correction can be written

\begin{equation}\label{pert2}
\Delta {\cal E}_{0}^{(2)} = 
\overline{\sum}\frac{(U_{\alpha,\beta}^{\gamma,\delta})^{2}}{\epsilon_{\gamma}
+\epsilon_{\delta}-\epsilon_{\alpha}-\epsilon_{\beta}}
\end{equation}    

\noindent $\overline{\sum}$ indicates that the sum is taken over occupied 
levels $\gamma$ and 
$\delta$ and unoccupied ones $\alpha$ and $\beta$ (taking into
account spin degeneracy), giving a total 
number of contributions $K \sim n^2(m-n)^{2}$. For the ground-state,
one has
$\epsilon_{\gamma}+\epsilon_{\delta} < \epsilon_{\alpha}+\epsilon_{\beta}$,
and each term in the sum (\ref{pert2}) is negative. The 
contributions from each scattering process therefore add coherently 
resulting in a strong reduction of the ground-state energy. The latter
can be estimated  for an equidistant spectrum
by replacing sums by integrals in (\ref{pert2}).
This gives an homogeneous polynomial of order three in $n$ and $m$, 
each term being multiplied by a logarithmic correction. In the dilute
limit $1 \ll n \ll m$ the $m^3$ and $m^2n$ terms drops out exactly so that the
dominant contribution to the second order correction in the ground-state 
energy reads

\begin{equation}
\Delta {\cal E}_{0}^{(2)} \approx -A (U^{2}/\Delta) n^{2} m\log(m)
\end{equation}

\noindent The prefactor $A$ can be estimated $A \approx 7$; details of the 
perturbative treatment are presented elsewhere \cite{pert}. 

\begin{figure}
\epsfxsize=3.1in
\epsfysize=2.36in
\epsffile{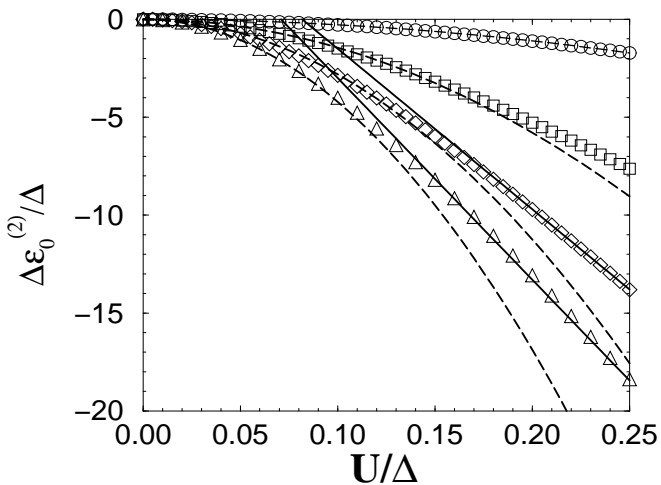}
\vspace{5mm}
\caption
{Average ground-state energies for 500 realizations 
of Hamiltonian (1) with $m=16$ orbitals, $n=2$ (circles), 4 (squares), 
6 (diamonds) and 8 (triangles) compared to the second order perturbative
result (3) (dashed lines) and to linear fits valid at large
$U$ (solid lines - see text).}

\end{figure}

In Fig.1 we
show the behavior of the ground-state energy for $n=2,$ 4, 6 and 8.
It is clearly seen that the average
ground-state energy exhibits the quadratic dependence
in $U$ predicted by (2-3) at weak $U$, while at larger $U$, this dependence 
becomes linear. Most importantly, this crossover 
occurs at a weaker $U$ for larger $n$. Both these 
features can be understood as one realizes that
for dominant fluctuations of interaction $U/\Delta \gg 1$
the Hamiltonian depends linearly in $U$. In this regime,
and for $n \gg 1$, the average many-body 
density of states is Gaussian with a
width $B$ well approximated by $B \sim n m U$ \cite{monfrench,newweiden}. 
This width thus gives a good
parametric estimate for the average ground-state energy in the asymptotic
regime $ {\cal E}_{0}^{(2)} \sim B \sim n m U$ \cite{spings} so that
equating the asymptotic and perturbative expressions for
the ground-state energy ($\Delta {\cal E}_{0}^{(2)} \sim nmU$)
gives a critical value 

\begin{equation}\label{uc}
U_{c} \sim \Delta/n
\end{equation}

\noindent The crossover between the two regimes occurs at a rate
given by $U_c$ and is inversely proportional to the number of particles.
As can be seen in Fig.1, below this threshold, the ground-state
energy is well approximated by the second order perturbation expression
(3), while above it, its $U$-dependence becomes linear {\it which cannot
be captured by any finite order of perturbation theory},
and we are forced to conclude that $U_c$ gives the radius of convergence for
perturbation theory in $U$. 

For
$U>U_c$, the ground-state starts to be mixed with higher excited states
and the number of Slater determinants $|J \rangle$
necessary to construct it becomes much
larger than one. A measure of this number
is given by the Participation Ratio (PR), which, for a given
many-body eigenstate $|\Psi_{\lambda}(U) \rangle$, is
defined as $\xi_{\lambda}=(\sum_{J} |\langle J |\Psi_{\lambda}(U) 
\rangle|^{2})^{2}/\sum_{J} |\langle J |\Psi_{\lambda}(U) \rangle|^{4}$.
If $|\Psi_{\lambda}(U) \rangle$ is made up of
one single Slater determinant (as it is for $U=0$), 
one has $\xi_{\lambda}=1$ whereas when
$|\Psi_{\lambda}(U) \rangle$ is homogeneously spread over
the $N$ Hilbert space basis states, $\xi_{\lambda} \sim N$.
To check the above threshold (\ref{uc}), we 
numerically calculated the PR for the ground-state wavefunction 
$|\Psi_{0}(U) \rangle$. Results obtained 
from the exact diagonalization of small systems of $n \le 8$ particles 
are shown in Fig.2.
We first note a faster increase of $\xi_{0}$ with $U$ at larger $n$ which
is not due to an increase of the Hilbert space size as it 
occurs before a significant portion
of the Slater determinant basis starts to be occupied. 
Most importantly, this increase is exponential
(note the vertical log-scale on Fig.2) : fixing e.g. $U/\Delta=0.2$,
one gets $\xi(n) \approx \exp(n/1.6)$. Since each order in perturbation theory
gives only an algebraic number of contributions, this exponential increase
of the PR cannot be captured by a finite, $n$-independent number of orders, 
and therefore indicates the divergence of the perturbation series 
as $n$ increases.
Another important feature of Fig.2 is the quite small value $U \ll \Delta$
at which $\xi_{0}$ starts to increase. This is quite surprising 
as $\Delta$ is the only energy scale close to the ground-state.
When the PR increases
beyond $\xi_{0} \ge 2$, second (and higher) order terms become more and more
important, until eventually perturbation theory breaks down. One can thus 
extract parametrically
the radius of convergence of perturbation theory from the PR
by defining a critical interaction strength $U_{c}$ from the condition
$\xi_{0}(U_{c})=2$. (This choice is arbitrary as long as one
chooses a threshold value 
$\xi_{0}(U_{c})$ much smaller than the total size of the Hilbert space 
as this would introduce an artificial $n$ and $m$ dependence in $U_{c}$.)
The inset to Fig.2 shows a $n-$ (and absence of $m-$) 
dependence of $U_{c}$ clearly confirming (\ref{uc}).

\begin{figure}
\epsfxsize=3.3in
\epsfysize=2.5in
\epsffile{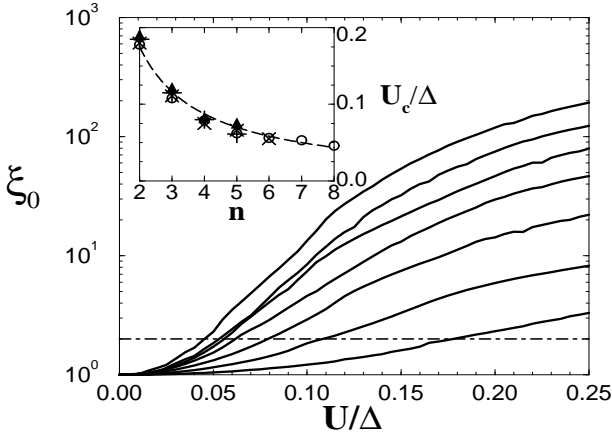}

\vspace{3 mm}

\caption
{Ground-state PR $\xi_0$ vs. $U/\Delta$ averaged over 
100 (for $n=8$) to 2000 (for 
$n=2$) realizations of 
Hamiltonian (1) for $m=16$ and $n$=2, 3, \ldots 8 (solid lines from 
bottom to top). The dotted dashed-line indicates the threshold 
$\xi_0=2$. Inset : Critical value of interaction 
$U_{c}/\Delta$ necessary to reach
$\xi_0=2$ vs. $n$, for $m$=10 (triangles), 12 (+),
14 ($\times$) and 16 (circles). The dashed line gives a decay 
$U_{c}/\Delta=0.35/n$.}
\label{fig:ipr}

\end{figure}

As yet another check of the threshold (\ref{uc}),
we consider the onset of level repulsion for
the spacing $s_1={\cal E}_1-{\cal E}_0$ between
the ground-state energy ${\cal E}_0$ and the first excited 
level ${\cal E}_1$. At $U=0$, this spacing
corresponds to a one-body excitation, and considering a randomly distributed
one-body spectrum, it is Poisson distributed 
$P_P(s_1)=\exp(-s_1)$. Switching on $U$ induces off-diagonal matrix
elements connecting the two lowest levels and leads to repulsion between
them. Making a two-level approximation incorrectly gives the onset
of this repulsion at $U = \Delta$. A correct treatment
goes as follows.
Most terms in the second order contribution (\ref{pert2}) are common 
to both the energies of the 
ground- and first excited states. The difference corresponds
to scattering from or onto the $(n/2)^{\rm th}$ and 
$(n/2+1)^{\rm th}$ orbitals, giving a contribution to the
energy difference between the lowest two levels

\begin{equation}\label{relf}
\overline{\sum} \frac{(U_{\alpha,n/2}^{\gamma,\delta})^2}
{\epsilon_{\alpha}+\epsilon_{n/2}-\epsilon_{\gamma}-\epsilon_{\delta}}-
\frac{(U_{\alpha,n/2+1}^{\gamma,\delta+1})^2}
{\epsilon_{\alpha}+\epsilon_{n/2+1}-\epsilon_{\gamma}-\epsilon_{\delta+1}}
\end{equation}

\noindent (There is an additional similar contribution arising from the
scattering from the $n/2$ and $n/2+1$ orbitals.)
We first make the substitution $\sigma((U_{\alpha,n/2}^{\gamma,\delta})^2-
(U_{\alpha,n/2+1}^{\gamma,\delta+1})^2) = U^2$
for IME fluctuations. Then, and 
considering an equidistant spacing, a simple counting shows that
the number of contributions in the above expression corresponding
to a given energy denominator $p \Delta$ grows as $p^2$ up to $p=n/2$ and 
decreases afterward. Neglecting logarithmic corrections,
one gets an estimate for the contributions to the relative 
fluctuations given by $\sim (U^2/\Delta) n^2$. This estimate can also be
obtained by replacing the sums in (\ref{relf}) with integrals.
Once these fluctuations become of the order of the spacing
between ${\cal E}_0$ and ${\cal E}_1$, $\sim (U^2/\Delta) n^2 \sim \Delta$, 
an avoided crossing occurs. This signals the
onset of level repulsion which therefore occurs at the rate given 
by (\ref{uc}).

\begin{figure}
\vspace{0.5cm}
\epsfxsize=3.3in
\epsfysize=2.5in
\epsffile{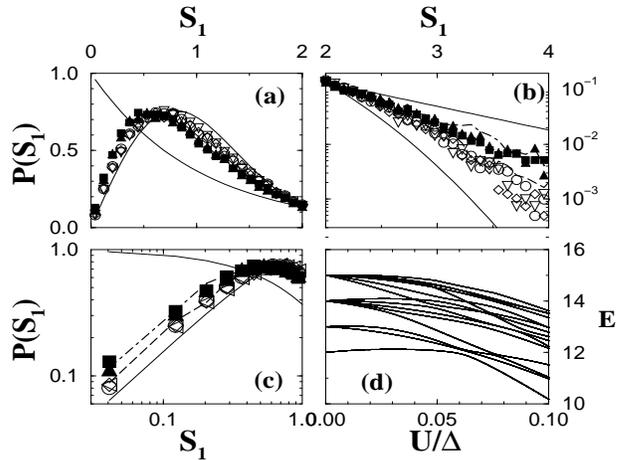}

\vspace{3 mm}

\caption
{(a,b,c) Distribution $P(s_1)$ 
of the first level spacing for 50000 realizations of 
Hamiltonian (1) with a random distribution of orbitals $\epsilon_{\alpha}$, 
$m=14$ (symbols) $n=3$, 4, \ldots 7, at an interaction 
strength $U/\Delta=1/n$. 
(a) Shows $P(s_1)$ on a normal scale, (b) its large $s_1$ tail on 
a semilogarithmic scale and (c) its small $s_1$ behavior on a log-log 
scale. Empty symbols correspond to odd $n$ and full symbols to even $n$.
Clearly, there is an even-odd dependence which is $m$-independent as 
shown by the superimposed values for $m=18$ and $n=3$ (dashed 
line) and $n=4$ (dotted-dashed line) on (b) and (c). On (a,b,c) the solid lines
show the Poisson and Wigner-Dyson distributions. 
(d) Shows one realization
of the spectrum for Hamiltonian (1) with equidistant one-body spectrum.
The spectrum correspond to the full 
sector $\sigma_{z}=0$ so that level crossings occur between levels with 
different $\sigma$.}
\label{fig:spectrum}

\end{figure}

This reasoning is confirmed on Fig.3 (a,b,c) where we present
the distribution $P(s_1)$ of the first level spacing for Hamiltonian (1) and
different number of particles $n$ at fixed interaction strength
$U=\Delta/n$. The two main features that can be seen are : 1) $P(s_1)$ at fixed
$U = U_c$ is $n$-independent, except that
2) $P(s_1)$ shows a clear even-odd dependence, in particular there is
more level repulsion for odd $n$. This second fact is understood
once one realizes that due to SRS, and within a 
given $(\sigma_z,\sigma)$-block, the first $U=0$ excitation is doubly
degenerate for $n$ odd, while it is not degenerate for $n$ even.
This results in a stronger effect of the interaction in the
former case and
the even-odd effect seen on Fig.3 (a,b,c). It is to be stressed that this is
purely a spin effect, this even-odd dependence does not appear for
spinless fermions. On Fig.3 (d) we also show a plot of the low-lying part
of the spectrum which 
shows an avoided crossing at a value $U/\Delta \approx 0.05$ much too small
to be explained by a two-level approximation. 
We stress that in (d) the avoided crossing is not accompanied by an 
increase of
the level spacing as would be the case in presence of non-fluctuating
IME, and which would delay the breakdown of perturbation theory.

Equation (\ref{uc}) predicts
a parametrically small radius of convergence for perturbation theory
beyond which the ground-state is composed of exponentially
many Slater determinants. It is somehow in opposition with earlier 
works \cite{sven} which, based on extrapolations from
higher excitation energies, predicted a border $\sim \Delta$.
The mechanism leading to the breakdown of perturbation theory at higher
excitation energies is however due to mixing between two-body levels which are
very close in energy, and thus give the dominant contribution in a second
order approach. In the immediate vicinity of the ground-state 
these processes with small denominators
do not exist as the lowest energy scale
is $\Delta$ - the approaches used in \cite{sven}
correspond then to a two-level approximation involving ground- and first
excited states. The presented theory 
show that the correct procedure is to retain a large number of 
small second order contributions which
have an enhanced effect close to the ground-state, 
as they acquire a well-defined sign and hence add coherently to give
the small threshold (\ref{uc}). One has to
conclude that the approach used in \cite{sven}, while fully
justified at a sufficient excitation energy, must break down somewhere
close to the ground-state. 

Random interaction models of the kind
investigated here describe interacting systems
restricted to energy differences of the order of the Thouless energy $E_c$
\cite{agkl}.
In disordered diffusive systems one has $E_c = g \Delta$
($g$ is the conductance) and introducing a cut-off at $E_c$ leads to
$m$, $n \sim g$. From (\ref{uc}) one would expect the breakdown 
of perturbation theory to occur for 
$U_c \sim \Delta/g$, precisely at the amplitude of fluctuations of IME
in these systems \cite{agkl,blanter}. The studied model
neglects however two ingredients which must be considered
when dealing with more realistic systems. First, correlations between
the two terms in (\ref{relf}) significantly reduce the value of
this latter expression below
the uncorrelated estimate $U^2 \sim (\Delta/g)^2$. 
Second, the mean-field charge-charge interaction
(that we voluntarily neglected), when incorporated self-consistently
into the one-body spectrum, leads to a strong reduction of the
one-body density of states \cite{kopietz}, hence an increase of the level spacing
at the Fermi level against which (\ref{relf}) must
be weighted, and an increase of the denominators for the infrared contributions
to second order perturbation theory. 
It is expectable that these two facts ensure the convergence of 
perturbation theory and the stability of the ground-state against off-diagonal
fluctuations in diffusive metallic systems. 

In summary, based on 
second order perturbation theory we predicted a fast reordering of
the ground-state for systems of randomly interacting fermions
which we confirmed numerically. While the influence of this
reordering on the quasiparticle excitations is at present unclear,
we found that it is
accompanied by a change of statistics of the first energy excitation above
the ground-state. Similar investigations
in lattice models where this onset of level repulsion
may influence the system's transport properties \cite{kohn}, and for which
correlations between IME should
play an important role (neglected in the presently studied model) 
are highly desirable.

Numerical computations were performed at the Swiss Center
for Scientific Computing. Work supported by the
Swiss National Science Foundation.


\begin{thebibliography}{99}

\vspace{-0.5cm}
    
\bibitem{altshklo} B.L. Altshuler and B.I. Shklovskii, Zh. Eksp. Teor.
Fiz. {\bf 91}, 220 (1986).

\bibitem{nuketh} R.A. Janik et. al., Phys. Rev. Lett. {\bf 81}
264 (1998); J.C. Osborn and J.J.M. Verbaarschot, Phys. Rev. Lett.
{\bf 81}, 268 (1998).

\bibitem{french} J. B. French and S.S.M. Wong, Phys. Lett. {\bf 33B},
447, (1970); {\bf 35B}, 5 (1971); O. Bohigas and J. Flores, Phys. Lett. 
{\bf 34B}, 261, (1971); {\bf 35B}, 383 (1971).

\bibitem{flam} V.V. Flambaum et. al., Phys. Rev. A {\bf 50}, 267 (1994);
V.V.Flambaum, F.M.Izrailev and G.Casati, 
Phys. Rev. E {\bf 54} (1996) 2136.

\bibitem{zelev} V. Zelevinsky et. al., 
Phys. Rep. {\bf 276}, 85 (1996).

\bibitem{agkl} B.L. Altshuler et. al.,
Phys. Rev. Lett. {\bf 78}, 2803 (1997).

\bibitem{mirlin} A.D. Mirlin and Y.V. Fyodorov, Phys. Rev. B {\bf 56},
13393 (1997); C. Mejia-Monasterio et. al., 
Phys. Rev. Lett. {\bf 81}, 5189 (1998);
X. Leyronas, J. Tworzydlo and C.W.J. Beenakker, 
Phys. Rev. Lett. {\bf 82},
4894 (1999).

\bibitem{yoram} Y. Alhassid, Ph. Jacquod and A. Wobst,
Phys. Rev. B {\bf 61}, R13357 (2000); Y. Alhassid and A. Wobst,
cond-mat/0003255.

\bibitem{spings} Ph. Jacquod and A.D. Stone, Phys. Rev. Lett. {\bf 84}, 
3938 (2000); Phys. Stat. Sol. {\bf 218}, 113 (2000).

\bibitem{pert} Ph. Jacquod and A.D. Stone, cond-mat/0102100.

\bibitem{nukespin} C.W. Johnson, G.F. Bertsch and D.J. Dean, Phys. Rev.
Lett. {\bf 80}, 2749 (1998); R. Bijker, A. Frank and S. Pittel, Phys.
Rev. C {\bf 60}, 021302 (1999); L. Kaplan and T. Papenbrock,
Phys. Rev. Lett. {\bf 84}, 4553 (2000); 
S. Drozdz and M. W\'ojcik, nucl-th/0007045;
D. Kusnezov,  Phys. Rev. Lett. {\bf 85}, 3773 (2000).

\bibitem{sven} S. {\AA}berg, Phys. Rev. Lett. {\bf 64}, 3119 (1990);
Ph. Jacquod and D. L. Shepelyansky, 
Phys. Rev. Lett. {\bf 79}, 1837, (1997);
P.G. Silvestrov, Phys. Rev. 
Lett. {\bf 79}, 3994 (1997).

\bibitem{monfrench} K.K. Mon and J.B. French, Ann. Phys. {\bf 95}, 90 (1975).

\bibitem{newweiden} L. Benet, T. Rupp and H.A. Weidenm\"uller,
cond-mat/0010426.

\bibitem{blanter} Ya.M. Blanter, Phys. Rev. B {\bf 54}, 12807 (1996).

\bibitem{kopietz} B.L. Altshuler and A.G. Aronov in ``{\it
Electron-electron Interaction in Disordered Systems}'', A.J. Efros and 
M. Pollak Eds, Elsevier (1985); P. Kopietz, Phys. Rev. Lett. {\bf 81}, 
2120 (1998); Ya. M. Blanter and M.E. Raikh, Phys. Rev. B {\bf 63}, 075304
(2001).

\bibitem{kohn} W. Kohn, Phys. Rev. A {\bf 133}, 171 (1964); R. Berkovits and Y.
Avishai, Phys. Rev. Lett. {\bf 76}, 291 (1996).

\end{thebibliography}
\end{document}